\pdfoutput=1

\documentclass[11pt,twoside,a4paper,cmspaper,final,collab]{cms-tdr}
\def\svnVersion{177388}\def\cmsCernNo{2012-264}\def\cmsCernDate{\today}\def\cmsMessage{Submitted to the Journal of High Energy Physics}

\begin{document}\cmsNoteHeader{TOP-12-017}

\hyphenation{had-ron-i-za-tion}
\hyphenation{cal-or-i-me-ter}
\hyphenation{de-vices}

\RCS$Revision: 154915 $
\RCS$HeadURL: svn+ssh://svn.cern.ch/reps/tdr2/papers/TOP-12-017/trunk/TOP-12-017.tex $
\RCS$Id: TOP-12-017.tex 154915 2012-10-26 15:52:52Z gerber $
\newlength\cmsFigWidth
\ifthenelse{\boolean{cms@external}}{\setlength\cmsFigWidth{0.85\columnwidth}}{\setlength\cmsFigWidth{0.4\textwidth}}
\ifthenelse{\boolean{cms@external}}{\providecommand{\cmsLeft}{top}}{\providecommand{\cmsLeft}{left}}
\ifthenelse{\boolean{cms@external}}{\providecommand{\cmsRight}{bottom}}{\providecommand{\cmsRight}{right}}
\renewcommand{\tt}{\ttbar\xspace}
\newcommand{\pdf}{\ensuremath{\mathrm{pdf}}}
\newcommand{\mzp}{\ensuremath{m_{\cPZpr}}}
\newcommand{\ptrel}{\ensuremath{p_{\text{T,rel}}}}
\newcommand{\htlep}{\ensuremath{L_{\text{T}}}}
\newcommand{\ptlep}{\ensuremath{p_{\text{T}}^{\text{lepton}}}}
\newcommand{\htall}{\ensuremath{H_{\text{T}}^{\text{all}}}}
\newcommand{\V}{\ensuremath{\mathrm{V}}}
\newcommand{\Zprime}{\cPZpr}
\newcommand{\pp}{\Pp\Pp}
\newcommand{\chisq}{\ensuremath{\chi^2}}
\newcommand{\Mttbar}{M_{\ttbar}}

\cmsNoteHeader{TOP-12-017} 

\title{Search for resonant \ttbar production in lepton+jets events in $\pp$ collisions at $\sqrt{s}=7\TeV$}

\author{The CMS Collaboration}

\date{\today}

\abstract{
A model-independent search for the production of heavy resonances decaying
into top-antitop quark pairs is presented. The search is based on events containing 
one lepton (muon or electron) and at least two jets selected from data samples 
corresponding to an integrated luminosity of 4.4--5.0\fbinv collected in $\pp$ collisions at $\sqrt{s}=7\TeV$.
Results are presented from the combination of two dedicated searches optimized for boosted production and production at threshold.
No excess of events is observed over the expected yield from the standard model processes.
Topcolor \cPZpr\ bosons with narrow (wide) width are excluded at 95\% confidence level for masses below $1.49~(2.04)\TeV$
and an upper limit of 0.3~(1.3)\unit{pb} or lower is set on the
production cross section times branching fraction
for resonance masses above  $1\TeV$. Kaluza--Klein excitations of a gluon with masses below $1.82\TeV$
(at 95\% confidence level) in the Randall--Sundrum model are also excluded, and an upper limit of $0.7\unit{pb}$
or lower is set on the production cross section times branching fraction for resonance masses above $1\TeV$.
}

\hypersetup{%
pdfauthor={CMS Collaboration},%
pdftitle={Search for resonant t t-bar production in lepton+jets events in pp collisions at sqrt(s)=7 TeV},
pdfsubject={CMS},%
pdfkeywords={CMS, physics, top quark}}

\maketitle 
\section{Introduction}
\label{intro}

The top quark is the heaviest known fermion, making it a powerful benchmark to extend our
understanding of the origin of mass. Because of its large mass, the top quark plays a central
role in several theories beyond the standard model (SM). These theories predict the
existence of heavy resonances that manifest themselves as an additional
resonant component to the SM \ttbar production.
Examples of such resonances, which decay preferentially into \ttbar,  include models with massive color-singlet
$\Z$-like bosons in extended gauge theories~\cite{zprime_Rosner,zprime_Lynch,zprime_Carena},
colorons~\cite{Hill1991419,Jain11124928,Hill:1993hs,Hill:1994hp} or axigluons~\cite{axigluon,Choudhury:2007ux}, models in which a
pseudoscalar Higgs boson may couple strongly to top quarks~\cite{pseudohiggs}, and models
with extra dimensions, such as
Kaluza--Klein (KK) excitations of gluons~\cite{Agashe:2006hk} or gravitons~\cite{graviton} in various extensions of the Randall--Sundrum model~\cite{RandallSundrum}.

Recent models~\cite{Bai:2011ed,Frampton2010294,PhysRevD.83.114027,PhysRevD.77.014003,Alvarez:2011hi}
aimed at explaining the $\ttbar$ charge asymmetry observed at the
Tevatron~\cite{PhysRevLett.101.202001, PhysRevLett.100.142002, PhysRevD.83.112003, PhysRevD.84.112005}
predict resonances in the 0.7--3\TeV mass range with production cross sections
of the order of a few pb and add renewed interest to the sub-\TeVns mass region.
Independent of the exact model, resonant \ttbar production could be visible in
the reconstructed invariant mass spectrum ($\Mttbar$).

Searches performed at the Tevatron have set upper limits on the production cross section of
narrow resonances ($\Zprime$ with mass below ${\sim}900\GeV$) decaying into
\ttbar~\cite{cdftt1,cdftt2,cdftt3,d0_resonance,cdf_resonance,d0_resonance_2008}.
Similarly, searches at the Large Hadron Collider (LHC) have set sub-pb limits on the production
cross section of resonances in the 1--3\TeV mass range~\cite{cms-allhad,atlas-ljets,atlas-ljets-boosted}.

In this paper, we present a model-independent search for the production of heavy resonances
decaying into \ttbar using data collected by the Compact Muon Solenoid (CMS)
experiment in $\pp$ collisions at $\sqrt{s} = 7\TeV$  at the LHC. Using samples
corresponding to an integrated luminosity of $4.4$--$5.0\fbinv$, we focus on the semileptonic
$\ttbar$ decay mode
$\ttbar \rightarrow (\PWp\cPqb)(\PWm\cPaqb) \rightarrow
{(\mathrm{q}_{1}\overline{\mathrm{q}}_{2}\cPqb) (\ell^-} \overline{\nu}_{\ell} {\cPaqb)}$ (or charge conjugate) wherein one
$\PW$ boson decays to an electron or muon and a neutrino, and the other decays hadronically.
The range of $0.5$--$3\TeV$ in $\Mttbar$ is covered by the combination
of two dedicated searches: one optimized for resonances with masses smaller than $1\TeV$ (threshold region), and a second one optimized for masses
larger than  $1\TeV$ (boosted region).
Both regions increase the sensitivity of the search by identifying jets
originating from the hadronization of b quarks (b jets), and separating the samples into various
categories depending on the lepton flavor, the number of jets, and the number of b jets.
The resulting samples are dominated by SM $\ttbar$ and $\PW$ bosons
produced in association with jets.
A limit on the production cross section of heavy resonances is extracted by performing
a template-based statistical evaluation of the reconstructed $\Mttbar$ distribution.

The CMS detector is briefly described in Section~\ref{sec:detector}. Section~\ref{sec:MC} provides details on the
data and simulated samples used in the analyses. Sections~\ref{sec:selection} and~\ref{sec:EventReco} describe the event selection and
the $\ttbar$ event reconstruction, respectively. The main sources of systematic uncertainty in the analyses are described
in Section~\ref{sec:systematics}. Results are shown in Section~\ref{sec:results} and a summary is provided in Section~\ref{sec:conclusions}.

\section{The CMS Detector}
\label{sec:detector}

The central feature of the CMS apparatus is a superconducting solenoid,
$13\unit{m}$ in length and $6\unit{m}$
in diameter, which provides an axial magnetic
field of $3.8\unit{T}$.  The bore of the solenoid is outfitted with various
particle detection systems.  Charged particle trajectories are
measured by the silicon pixel and strip trackers, covering $0 < \phi <
2\pi$ in azimuth and $|\eta |<2.5$, where pseudorapidity $\eta$ is
defined as $\eta =-\ln[\tan{(\theta/2)}]$, with $\theta$ being the
polar angle of the trajectory of the particle with respect to the
counterclockwise beam direction.
A crystal electromagnetic calorimeter
and a brass/scintillator hadronic calorimeter surround
the tracking volume.
In this analysis the calorimetry provides high-resolution energy and
direction measurements of electrons and hadronic jets.
Muons are measured in
gas-ionization detectors embedded in the steel return yoke outside the solenoid.
The detector is nearly hermetic, allowing for momentum balance
measurements in the plane transverse to the beam directions, which are used to infer the presence of neutrinos in events.
A two-tier trigger system selects the most interesting $\Pp\Pp$ collision
events for use in physics analysis.
A more detailed description of
the CMS detector can be found in Ref.~\cite{JINST}.

\section{Data and Simulated Samples}
\label{sec:MC}

The data analyzed for the threshold analyses were recorded with
triggers requiring a single isolated (defined in Section~\ref{subsec:ThresholdSelection}) muon or electron
with a transverse momentum ($\pt$) threshold of $17\GeV$ or $25\GeV$, respectively, in combination
with a number of jets with a  $\pt$ threshold of $30\GeV$.
Events containing an electron were required to have three or
more jets throughout the data-taking period, while the minimum number of
jets in events containing a muon increased from zero to three as the instantaneous
luminosity increased.
The data analyzed for the boosted analyses were recorded with triggers requiring
one muon with a $\pt$ threshold of $40\GeV$ or one electron  with
a $\pt$ threshold of $65\GeV$, with no isolation requirements on the leptons.
To avoid too high a trigger rate, the electron trigger was prescaled for the highest instantaneous luminosities. This resulted
in a loss of $0.6 \fbinv$ of integrated luminosity for the boosted electron analysis compared to the other channels.
No additional requirements were made on the jets or missing transverse energy in the triggers used for the
boosted analyses.

Offline, we use a particle-flow~\cite{PFT-09-001} based event reconstruction,
which combines information from each subdetector, including charged particle tracks from the
tracking system and deposited energy from the electromagnetic and hadronic calorimeters, to
reconstruct all particles in the event. Particles are classified as electrons, muons,
photons, charged hadrons, and neutral hadrons. Particles identified as originating from
multiple primary collisions at high instantaneous luminosity (pileup) are removed from the event.

Muons are reconstructed using the information from the muon chambers
and the tracking detectors \cite{MUO-10-002}.
Tracks are required to have at least 11 hits including
at least one in the pixel layers.
The tracks must also pass within
$0.02\unit{cm}$ of the beam spot in the plane
transverse to the beam, and within $1\unit{cm}$ along the beam axis.

Electron candidates are initially identified by matching a track to a cluster of
energy in the electromagnetic calorimeter. Candidates are selected~\cite{EGM-10-004} using shower-shape
information, the quality of the track and the spatial match between the track and
electromagnetic cluster, the fraction of total cluster energy in the
hadronic calorimeter, and the amount of activity in the surrounding
regions of the tracker and calorimeters.
Electrons coming from photon
conversions in the detector material are rejected if there are missing
hits in the inner tracker layers or if there is another close track with opposite charge and
with a similar polar angle.

Jets are reconstructed by clustering the particle-flow candidates not identified as leptons
using an anti-\kt\ algorithm with a
distance parameter $R=0.5$~\cite{Cacciari:2008gp}. Corrections are applied to
account for the dependence of the detector response to jets as a function of $\eta$
and \pt~\cite{JETJINST} and the effects of pileup.
The jets associated to b quarks are identified using an algorithm that
reconstructs the secondary vertex corresponding to the decay of a B hadron.
When no secondary vertex is found, the significance of the impact parameter  with respect to the primary vertex
of the second most displaced track is used as a discriminator to
distinguish decay products of a B hadron from prompt tracks~\cite{CMSbAlgo}.

The negative of the vector sum of the momenta of all reconstructed particles in the plane transverse to the beam
is the missing transverse momentum~\cite{METJINST}, with
magnitude denoted by missing transverse energy \MET.

The SM background processes are simulated by \MADGRAPH 5.1.1~\cite{MadGraph},
\PYTHIA 6.4.24~\cite{pythia}, and \POWHEG~\cite{powheg}
event generators using \rm{CTEQ6L} parton distribution functions of
the proton~\cite{cteq}.
The generated events are subsequently
processed with \PYTHIA to provide the showering of the partons and fully simulated with CMS
software based on \GEANTfour~\cite{Agostinelli2003250,1610988}.

The $\PW$ boson and Drell--Yan production in association with up to four jets are simulated with \MADGRAPH,
with additional jet
production described via matrix elements matched to parton showers using the MLM prescription~\cite{mlm}
 with a matching threshold of $20\GeV$. The next-to-next-to-leading order (NNLO) production cross sections
 times branching fractions into leptons (electrons, muons and taus)
are used~\cite{fewz}: $31.3\unit{nb}$ for $\PW$, and $3.05\unit{nb}$
for Drell-Yan production of dilepton final states with invariant mass $>50\GeV$.
The background from Drell-Yan production of dilepton final states with invariant mass
$<50\GeV$ is negligible. The contribution from QCD multijet processes is obtained directly from 
data as described in Section~\ref{sec:EventReco}.

The SM \tt events are generated with \MADGRAPH, assuming a top-quark mass of $172.5\GeV$.
Higher-order gluon and quark production is described by the matrix elements with up to three extra partons
beyond the \tt system. The chosen threshold for the matching is
$40\GeV$, which ensures a smooth transition from the matrix element to the parton showering description.
An additional \tt sample is generated using \textsc{powheg}
to provide a cross-check and to estimate systematic uncertainties in the modeling.
The inclusive \tt cross
section value of $157.5\unit{pb}$ is used~\cite{mcfm:tt1,mcfm:tt2}.

Single top-quark production is modeled in \textsc{powheg}.
The approximate NNLO cross sections of $42\unit{pb}$ and $3.2\unit{pb}$
are used for $t$-channel and $s$-channel single top-quark production, respectively, along with the corresponding
single \cPaqt-quark production cross sections of $23\unit{pb}$ and $1.4\unit{pb}$.
The approximate NNLO value of $7.9\unit{pb}$
is used for Wt and
$\PW\cPaqt$ associated production~\cite{singleTop-s,singleTop-t,singleTop-tW}.

Finally, as reference models for new physics, we use the sequential standard model (SSM) topcolor \zp bosons with
a natural width
$\Gamma_{\Zprime}$ equal to 1.2\% (narrow width) and 10\% of the $\Zprime$ mass $m_{\Zprime}$ based on~\cite{Hill1991419,Jain11124928,Hill:1993hs,Hill:1994hp}
and KK gluons based on~\cite{Agashe:2006hk}. Signal samples are generated with \PYTHIA 8.145 
with a range of masses between $500\GeV$ and $3\TeV$. Only decays into \tt are simulated in the \zp samples.
The KK gluons are simulated with branching fractions to \tt of
0.93, 0.92, 0.90, and 0.87 for resonance masses of 1, 1.5, 2 and 3\TeV.

\section{Event Selection}
\label{sec:selection}

To study the range of $0.5$--$3\TeV$ in $\Mttbar$, two complementary strategies are pursued: firstly,
the threshold search focuses on the $0.5$--$1\TeV$ mass range
using criteria optimized to identify top quarks produced
with a small boost in the detector frame and hence with well-separated decay products.
In this region, if all decay products are reconstructed within the kinematic acceptance, we expect the final state to contain
exactly one isolated lepton, four jets produced by the four quarks (two of which are b jets) in the semileptonic \tt decay, and \MET.

\begin{figure}
\begin{center}
\includegraphics[width=0.48\textwidth]{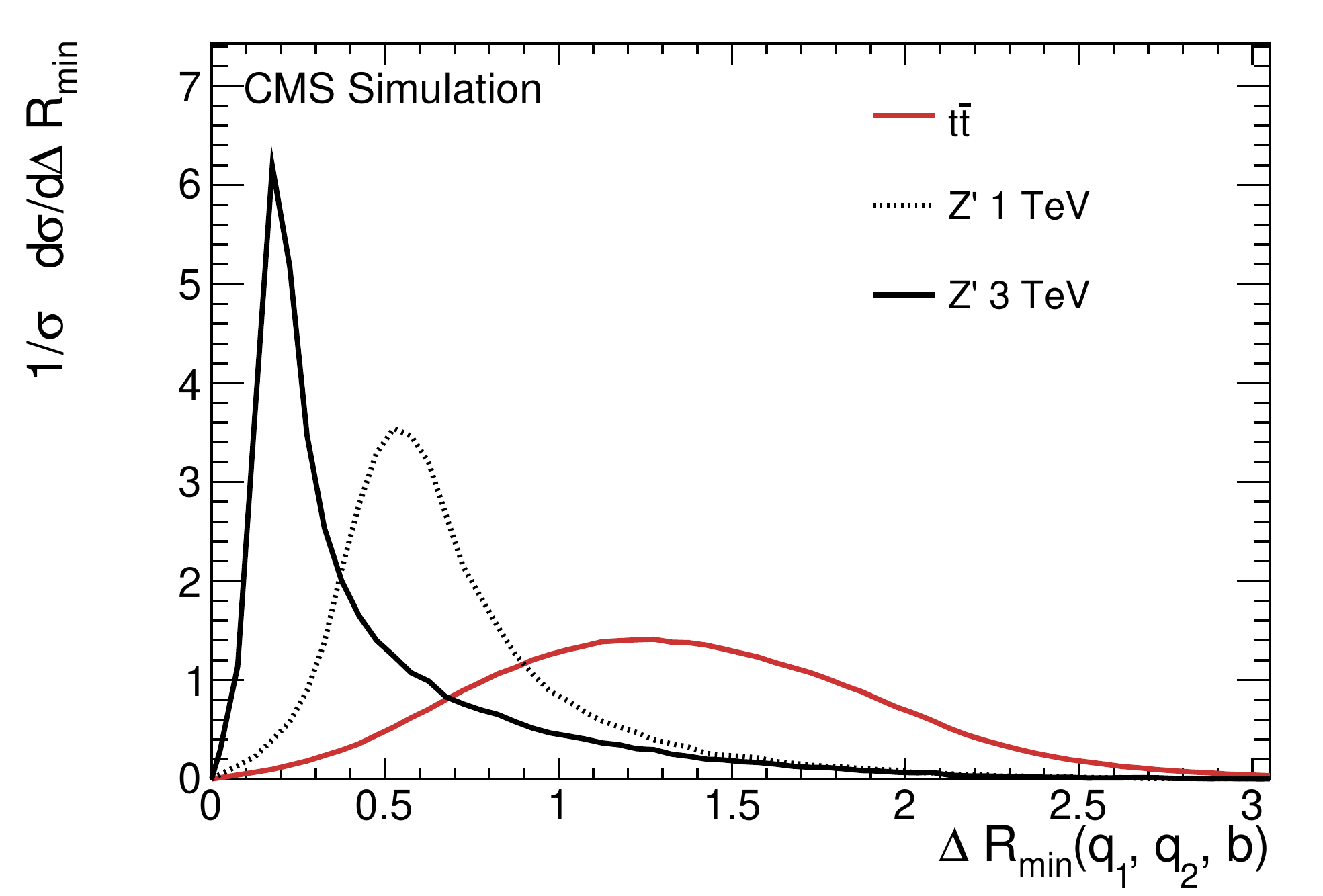}
\end{center}
\caption{The distribution of the minimum $\Delta R$ of all three possible pairings between
the three quarks $(\mathrm{q}_1, \mathrm{q}_2, \cPqb)$ of the hadronic top-quark decay for SM \tt production and two different $\Zprime$ mass hypotheses.
For events with $\Delta R_{\text{min}}$ smaller than the parameter $R=0.5$ in the jet clustering, jets merge and fewer than three
jets are reconstructed.}
\label{fig:gendr}
\end{figure}

Secondly, for resonance masses above $1\TeV$, the highly Lorentz-boosted top quarks will yield collimated
decay products that are partially or fully merged.
This can be seen in Fig.~\ref{fig:gendr}, which shows that in the boosted region the angular distance between
the partons is smaller than the jet clustering distance parameter.
As a consequence, the products of the hadronically decaying top quark might be reconstructed as
fewer than three jets, and the leptons might not be isolated.
The boosted search thus selects events containing
one electron or muon with no isolation requirement and at least two jets.

\subsection{Threshold analyses}
\label{subsec:ThresholdSelection}

We select events containing either one isolated muon
with $\pt > 20\GeV$ and $|\eta| < 2.1$, or one isolated electron with
$\pt > 30\GeV$ and $|\eta| < 2.5$.
The isolation requirement is based on the ratio of the total transverse energy
observed from all hadrons and photons in a cone of size
$\Delta R =\sqrt{\left ( \Delta\phi \right )^2+\left (\Delta\eta \right )^2}< 0.4$ around the
lepton direction to the transverse momentum of
the lepton itself. This quantity is required to be
less than 0.125 for muons and less than 0.1 for electrons.
Events with two isolated lepton candidates are vetoed to reduce the background from Drell-Yan
and \tt production in which both $\PW$ bosons decay leptonically.

Events are further required to contain at least three jets with
$|\eta| < 2.4$ and $\pt>50\GeV$, and additional jets with  $|\eta| < 2.4$ and $\pt>30\GeV$, if any.
To enhance the rejection of background from $\PW$-boson and Drell-Yan
production in association with relatively low-\pt jets,
the leading jet is required to have $\pt> 70\GeV$.
Multijet background is suppressed further by requiring $\MET> 20\GeV$.
The fraction of simulated semileptonic signal events passing this selection varies from
16 to 35\% for resonance masses below $1\TeV$.

Events are then separated into eight categories
according to the lepton flavor (electron or muon), the number of jets, and the number of b-tagged jets. The categories
defined by jets are: events with three jets, of which at least one is b tagged;
events with four or more jets, of which none is b tagged; events with four or
more jets, of which exactly one is b tagged; and events with four or more jets, of which at least two are b tagged.

\subsection{Boosted analyses}
\label{subsec:BoostedSelection}

We select events containing either one muon  with $\pt > 42\GeV$ and $|\eta| < 2.1,$ or
one electron with $\pt > 70\GeV$ and $|\eta|<2.5$, and at least two jets with
$|\eta| < 2.4$ and $\pt > 50\GeV$. The leading jet $\pt$ lower
threshold is set to
$250\GeV$ ($150\GeV$) in the muon (electron) channel.
No isolation requirement is applied to the leptons.
Multijet background is reduced with a requirement on the $\Delta R$ separation in the 2D plane:
$\Delta R (\text{lepton}, \text{closest jet}) > 0.5$ or
$p_T^\text{rel}(\text{lepton}, \text{closest jet}) > 25 \GeV$.
Here, $p_T^\text{rel}$ is defined as the magnitude of the lepton momentum orthogonal to the closest jet axis,
where any jet with $\pt > 25\GeV$ is considered.
We also require the scalar quantity $\htlep > 150 \GeV$,
where $\htlep = \MET + \ptlep$.

In the electron channel only, the multijet background is further reduced by requiring that
$\MET > 50\GeV$ and applying a
series of topological requirements that ensure the missing transverse momentum
does not point along the transverse direction of the electron (e) or of the leading
jet (j):
\[-\frac{1.5}{75\GeV} \MET +1.5 < \Delta\phi\{({\Pe\; \text{or}\; j)}, \MET\} < \frac{1.5}{75\GeV} \MET +1.5. \]

Even though the lepton $\pt$ requirements are dictated by the trigger threshold, the leading jet
$\pt$ requirement is chosen so that the total transverse energy of the event (including $\MET$) is as close as
possible in both channels. In addition, we ensure the two channels contain no overlap with each other by vetoing events that contain a second lepton.

Events are separated into four categories
according to the lepton flavor (electron or muon) and the number of b-tagged jets: either
no b-tagged jets, or at least one b-tagged jet.
The fraction of simulated semileptonic signal events passing this selection varies from
13 to 24\% for resonance masses between 1 and $3\TeV$.

\section{The \tt Event Reconstruction}
\label{sec:EventReco}

The four-vectors of the top quark and antiquark candidates
are reconstructed by assigning the final state objects in each event to either the
leptonic or the hadronic leg of the $\tt$ pair decay. We then choose between the possible
hypotheses using the criteria described below that depend on the number of reconstructed jets.
This $\tt$ reconstruction process results in a unique value for the reconstructed $\Mttbar$ for each event.

First, the charged lepton and the $\MET$ are assigned to the leptonic leg, where
$\MET$ is interpreted as the transverse component of the momentum of the neutrino.
Imposing the condition that the invariant mass of the
lepton and neutrino is equal to the mass of the $\PW$ boson (80.4\GeV) allows the
construction of a quadratic equation for the longitudinal component of the momentum of
the neutrino.  In the absence of a real solution, the boosted analyses retain the real part of the
complex solution. The threshold analyses modify the components of \MET
by the minimal amount in $|\Delta {\MET}_x|+|\Delta{\MET}_y|$
to give one real solution, which results in an improved mass resolution.  If there are
two real solutions, hypotheses are built for both cases, effectively doubling the number of combinations for that event.

For events with four or more jets in the threshold analyses,
the choices of neutrino solution and jet association are made
simultaneously by forming a $\chi^2$
from the sum of the normalized squared deviations of the
leptonic top-quark mass, hadronic top-quark mass, hadronic $\PW$ mass, $\pt$ of the \tt system,
and the ratio of the  $\pt$ of the four selected jets to the $\pt$ of all jets in the event.
The central values and widths used are obtained from the
distributions of these quantities in the Monte Carlo simulation.
The $\chi^2$ is calculated for each possible combination, including the two neutrino
solutions if they are both physical. The b-tagged jets may only be associated to a b quark in the decay chain, thereby reducing the
number of possible combinations.
For each event, the combination with the smallest value of $\chi^2$ is chosen.
The association of jets to the $\PW$ boson and the b quarks is found to be
correct in approximately 80\% of simulated $\ttbar$ events for which the four jets in the decay chain are reconstructed.

For events with only three jets in the threshold analyses, it is assumed that two
jets from the $\ttbar$ decay may have merged. The leptonic $\PW$ boson
is first reconstructed
as described above. The solution for the longitudinal
neutrino momentum is chosen to give the closest match to the
leptonic top-quark mass when the
leptonic $\PW$ boson is combined with any of the three jets. The invariant
mass of the leptonic $\PW$ and all three jets together is then taken as an
estimate of $\Mttbar$.

For the boosted analyses, we allow for collimated
decay products that are partially or fully merged by
considering all hypotheses that have exactly one jet assigned to the leptonic leg,
and at least one jet assigned to the hadronic leg. A two-term $\chi^2$ is constructed
from the sum of the normalized squared deviations of the
leptonic top-quark mass and the hadronic top-quark mass.
For each event, the combination with the smallest value of $\chi^2$ (labeled $\chi^2_\text{min}$) is chosen.
Next, the event selection described in Section~\ref{subsec:BoostedSelection} is extended
by applying additional conditions that improve the overall sensitivity of the boosted analyses.
For the electron channel only, the transverse momentum of the reconstructed leptonic top quark is required
to be greater than $100\GeV$. We require $\chi^2_\text{min} < 8$ for both channels.
This value is chosen such that the efficiency for this cut is 50\% for signal and approximately 10\% for the $\PW$+jets background.
Finally, we categorize events according to the number of b-tagged jets as either
with no b-tagged jets, or with at least one b-tagged jet.

The multijet background contribution to each channel in the threshold analyses is determined from data.  A multijet-dominated sample is defined by removing the \MET requirement and selecting events containing fake leptons, defined as muon candidates with isolation values between 0.2 and 0.5, and electron candidates consistent with photon conversions. This sample is used to define templates for multijet background distributions used in the analyses, including the shape of the $\Mttbar$ distribution; templates for other SM backgrounds are taken from simulation. These templates are used to find the number of multijet events from a fit to the lepton $\eta$ (in the electron channel) or the $\pt$ of the vector sum of jet momenta (in the muon channel) in a sample that contains events that pass the selection cuts but have $\MET<20~\GeV$. The number of multijet events in the final sample is obtained by extrapolating the result to the $\MET>20~\GeV$ region using the normalization determined
in the sample with $\MET<20~\GeV$. In the boosted analyses the multijet contamination after the final selection is found to be negligible.

\begin{figure}
\centering
 \includegraphics[width=0.48\textwidth]{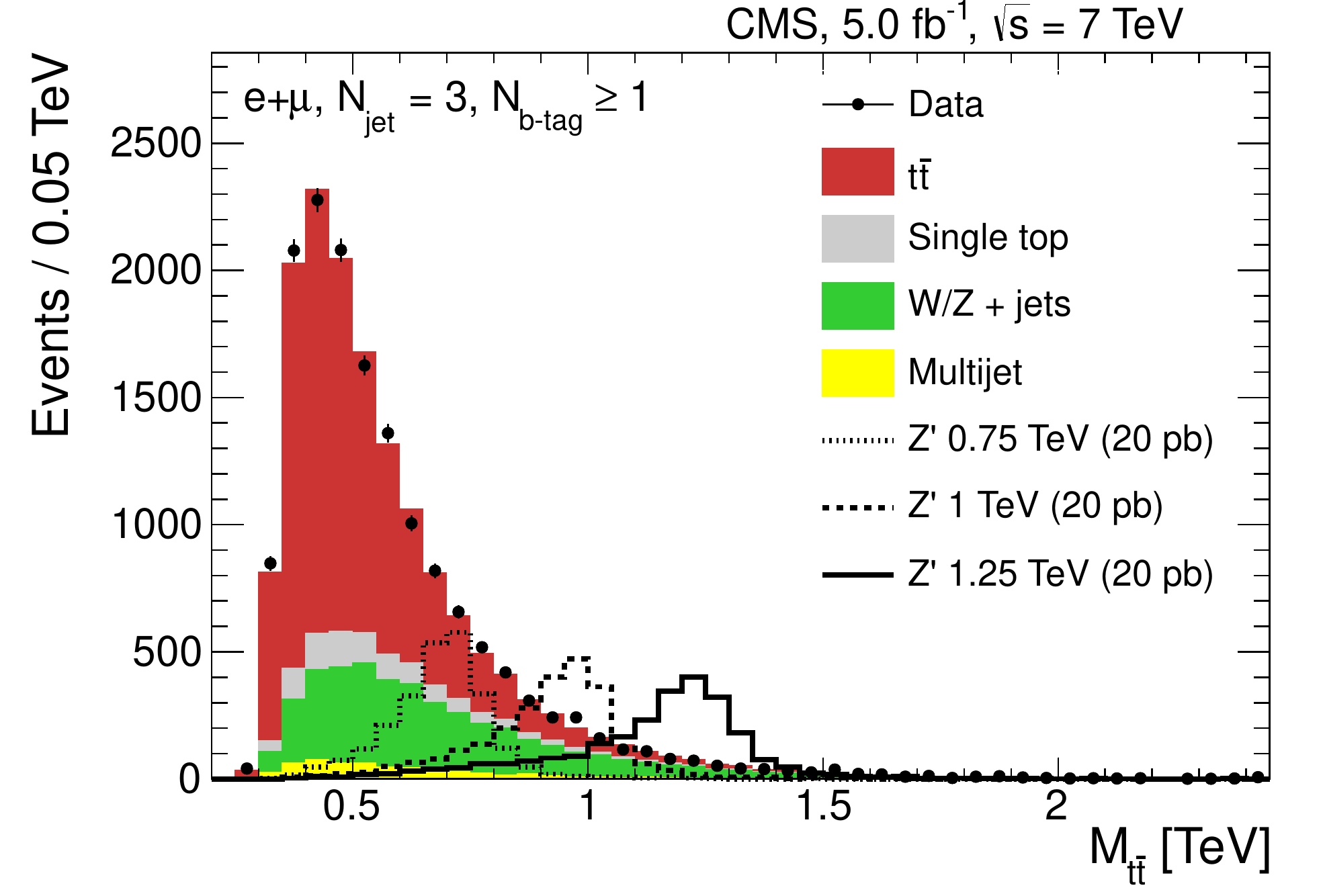}
 \includegraphics[width=0.48\textwidth]{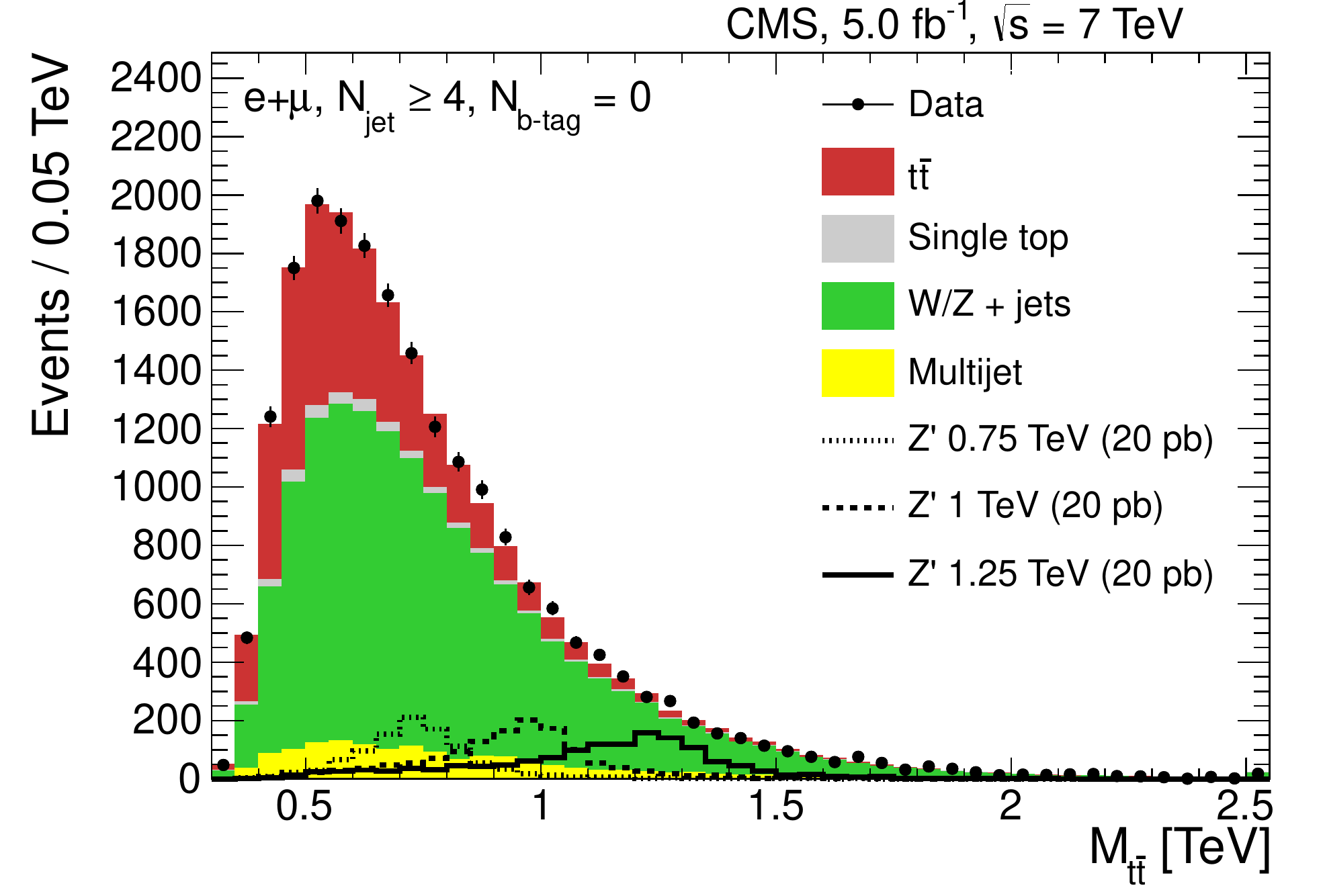}
 \includegraphics[width=0.48\textwidth]{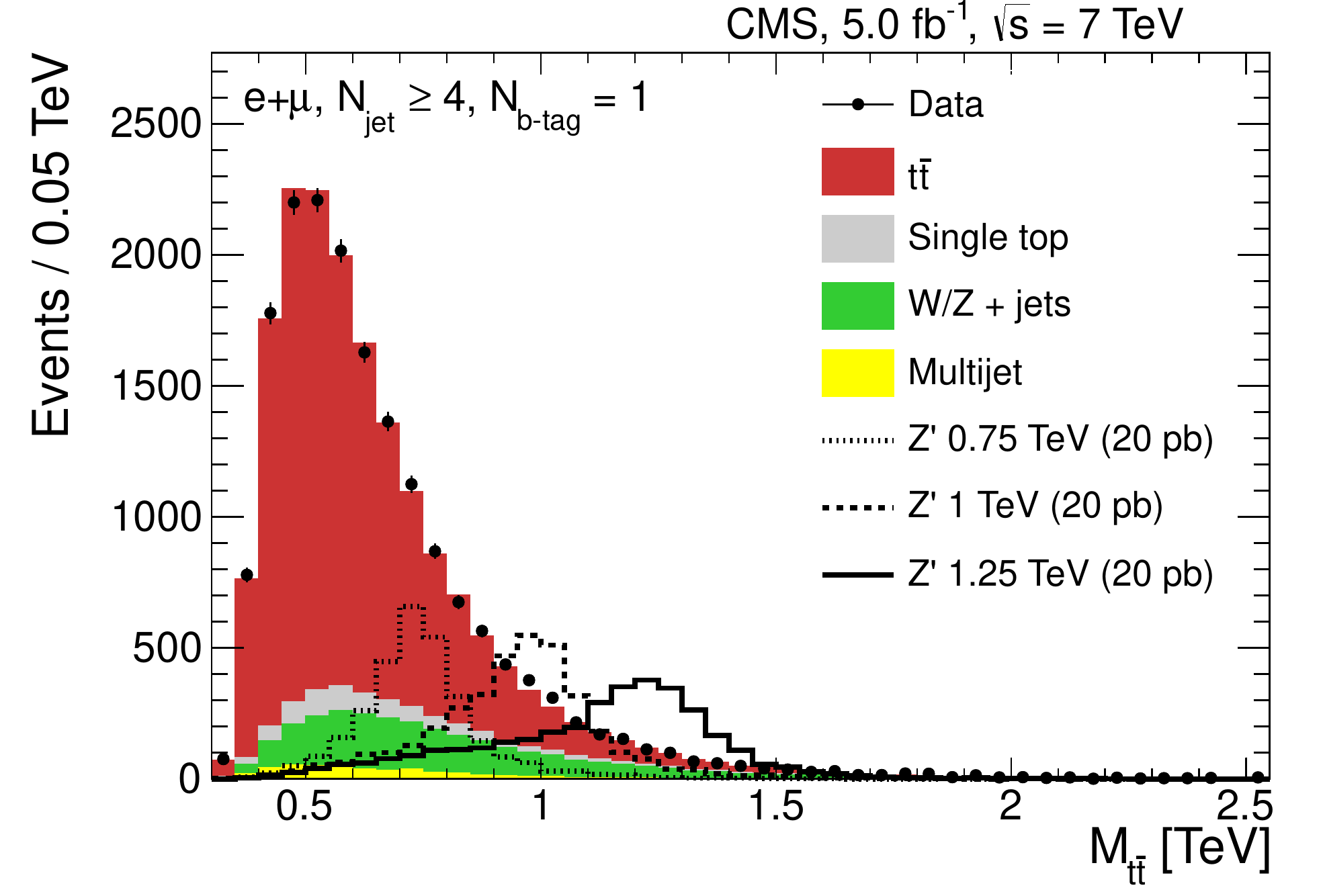}
 \includegraphics[width=0.48\textwidth]{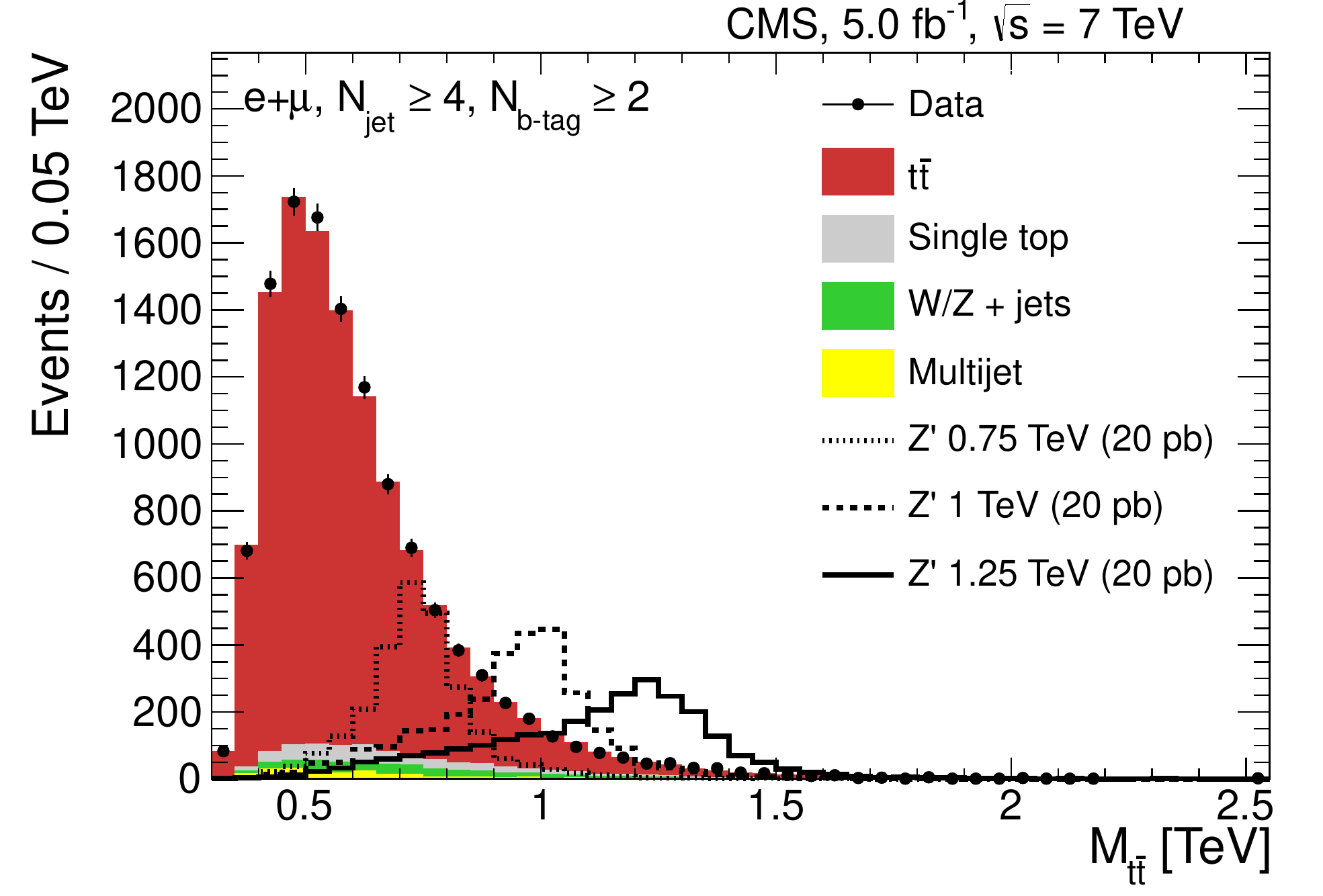}
\caption{Comparison of the reconstructed $\Mttbar$ in data and SM predictions for  the threshold analysis with
(a) 3 jets of which ${\ge}1$ b tagged, (b) 4 jets, none of which is b tagged,
(c)  4 jets of which one is b tagged, (d) 4 jets of which ${\ge}2$ are b tagged.
Expected signal contributions for narrow-width topcolor $\zp$ models at different masses are also shown.
For clarity, a cross section times branching fraction of 20\unit{pb} is used for the normalization of the
$\zp$ samples.
\label{fig:threshold-mttbar}}
\end{figure}

\begin{figure}
\centering
 \includegraphics[width=0.48\textwidth]{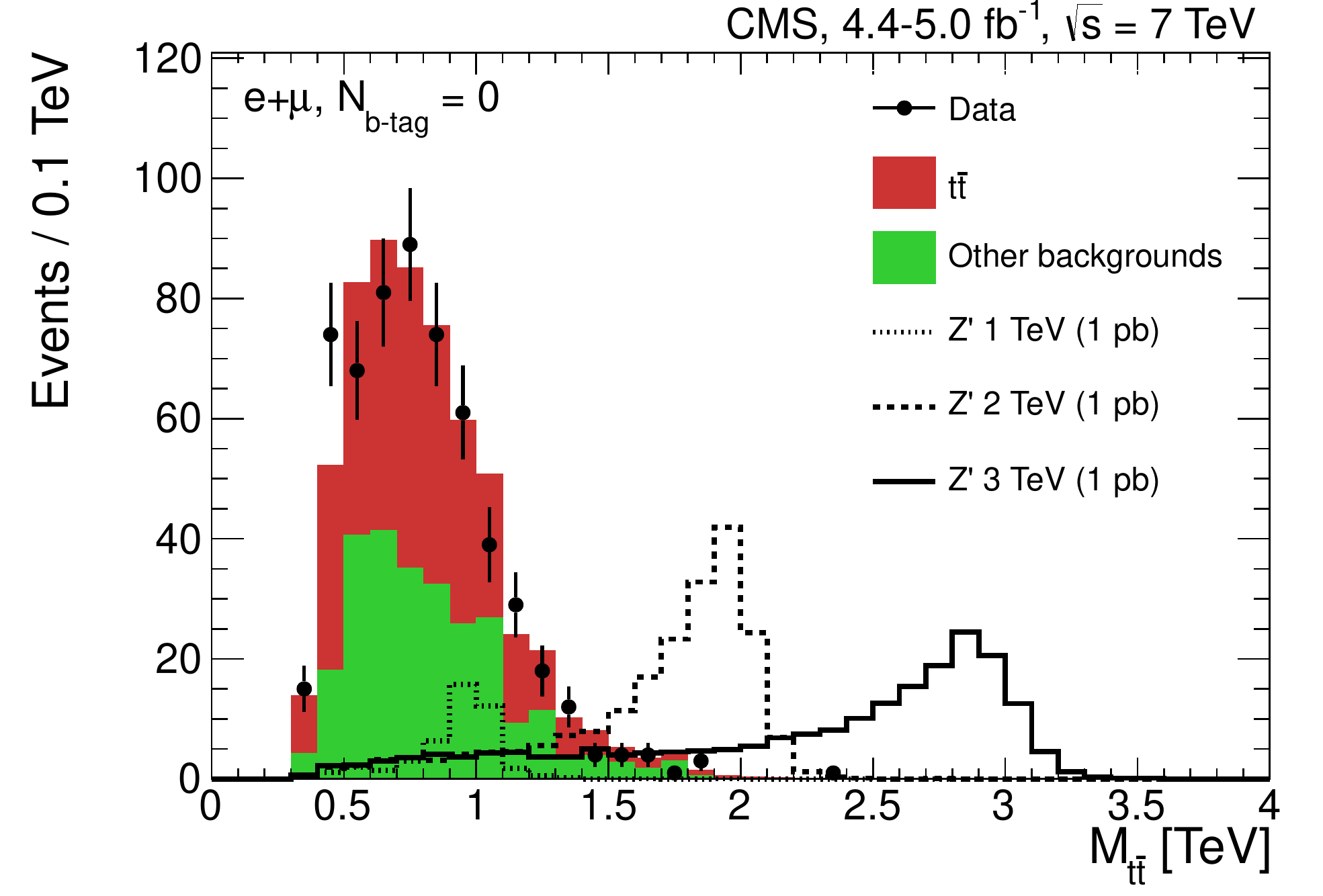}
 \includegraphics[width=0.48\textwidth]{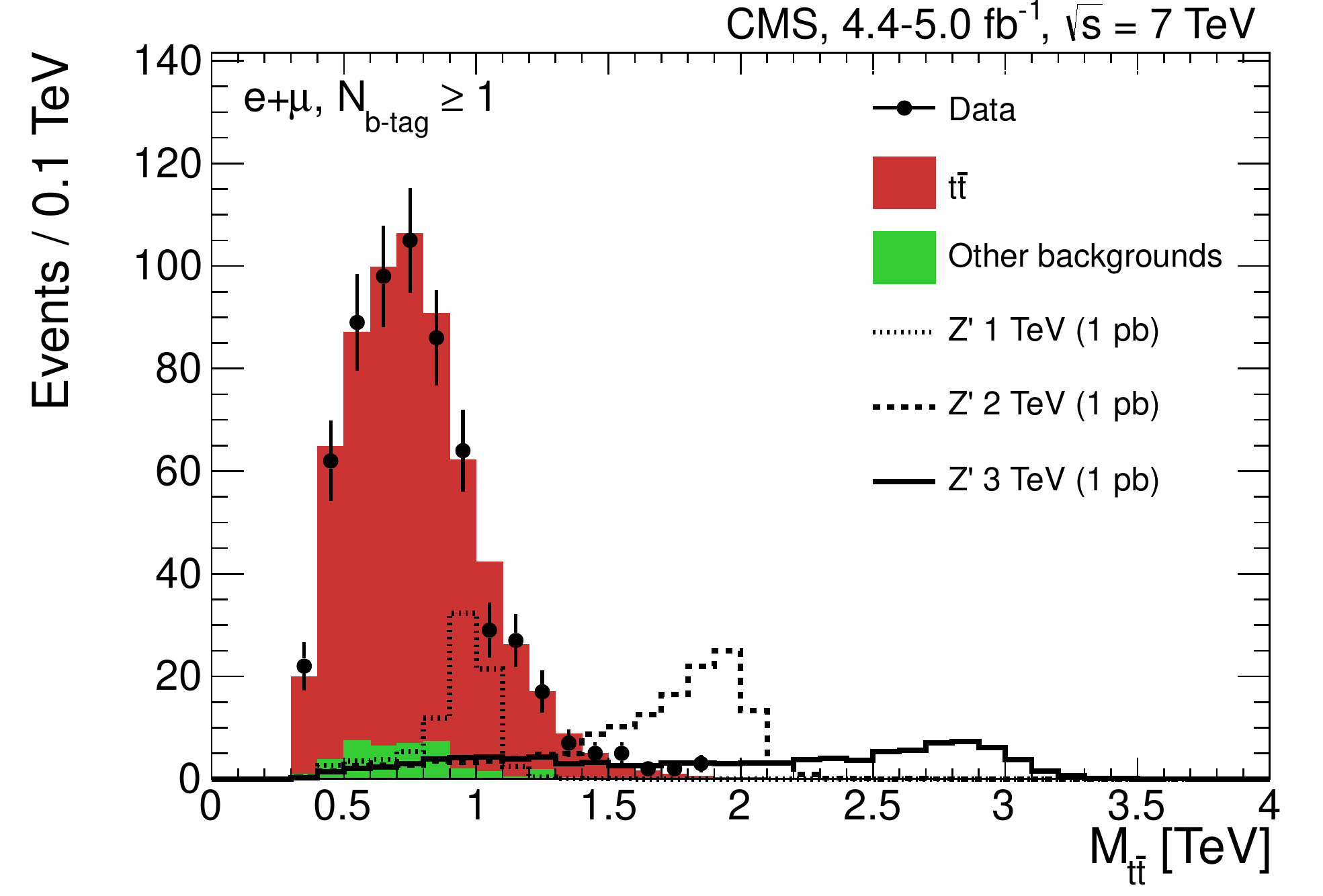}
 \includegraphics[width=0.48\textwidth]{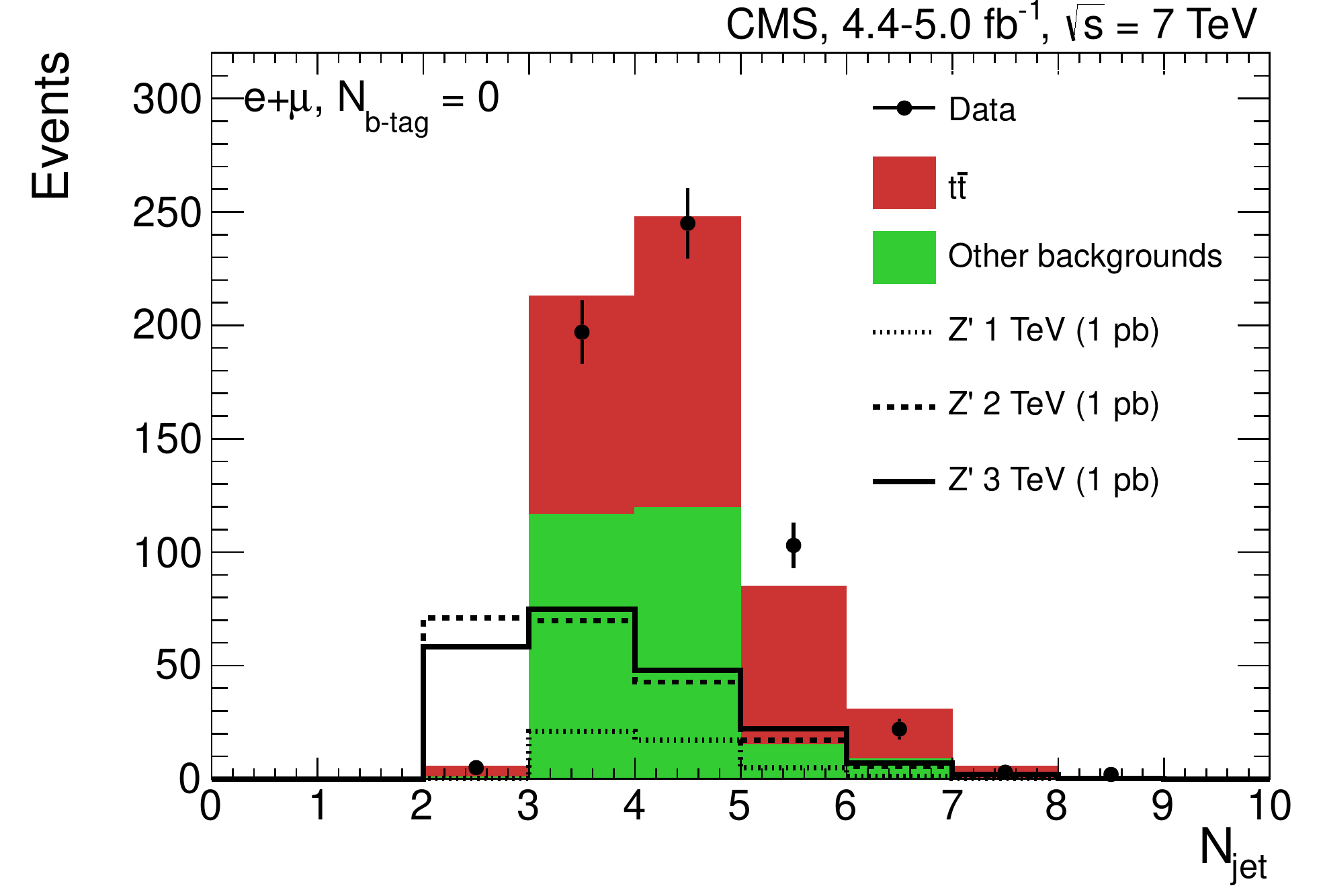}
 \includegraphics[width=0.48\textwidth]{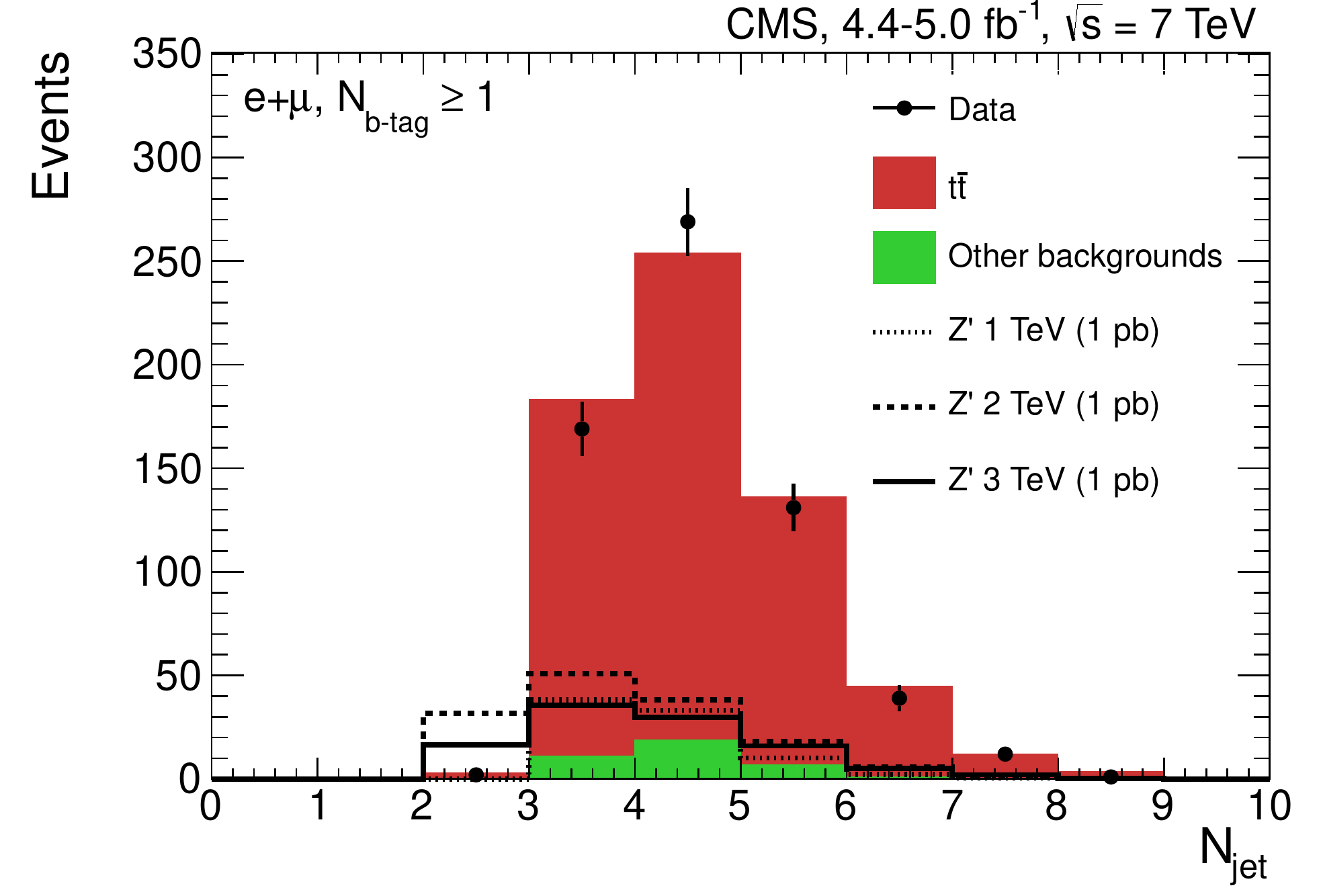}
\caption{Comparison of the reconstructed $\Mttbar$ in data and SM predictions for  the boosted analysis with
(a)  no b-tagged jets, (b) ${\ge}1$ b-tagged jets.
Comparison of the jet multiplicity distribution in data
and SM background predictions for the boosted analysis with
(c) no b-tagged jets, (d)  ${\ge}1$ b-tagged jets.
Expected signal contributions for narrow-width topcolor $\zp$ models at different masses are also shown.
A cross section times branching fraction of 1.0\unit{pb} is used for the normalization of the
$\zp$ samples.
\label{fig:boosted-mttbar}}
\end{figure}

The numbers of expected and observed events in each analysis channel for the threshold and boosted analyses
are summarized in Tables~\ref{tab:eventyieldthreshold} and~\ref{tab:eventyieldboosted}, respectively.
The $\Zprime$ samples are normalized arbitrarily to cross sections times branching fractions of 1\unit{pb}.
For the threshold analyses, the simulated samples are normalized to theoretical predictions.
For the boosted analyses, the yields of the simulated samples are normalized to data using scale factors
derived in a maximum likelihood fit to the $M_{\ttbar}$ distribution in both channels simultaneously.
This is done to allow for possible shortcomings of the theoretical predictions in the more extreme region of phase space probed by these channels. 
The likelihood is defined as described in 
Section 7, where the simulated samples are initially normalized to the theoretical predictions, 
but the normalization is allowed to vary within the uncertainties during the fitting procedure. 
Figures~\ref{fig:threshold-mttbar}~and~\ref{fig:boosted-mttbar} show the $M_{\ttbar}$ distributions for the threshold and boosted analyses,
respectively. Figure~\ref{fig:boosted-mttbar} also shows the distribution of the number of jets in the events for the boosted
analyses. It can be observed that, in the boosted region, the signal populates the 2-jet bin while the SM
background has larger jet multiplicity.
Good agreement is observed in all cases between data and the SM predictions.

\begin{table}
\topcaption{\label{tab:eventyieldthreshold} Number of expected and observed events in the threshold analyses for
an integrated luminosity of 5.0\fbinv.
The narrow-width $\Zprime$ samples are normalized to cross sections times branching fractions of 1\unit{pb}.
The other simulated samples are normalized to theoretical predictions.
The uncertainty in the total background corresponds to yield changes originating from the
systematic uncertainties associated with the jet energy corrections, jet energy resolutions, b tagging, and pileup.
The normalization uncertainties on the theoretical production cross sections are summarized in 
Section 6, and are not included in the quoted value. The statistical uncertainties for the simulated samples are negligible.}
 \begin{center}
    \begin{tabular}{|lrrrr|} \hline
\multicolumn{5}{|c|}{Threshold analyses, muon channel}\\ \hline
Sample & $N_{\text{jet}}=3$& $N_{\text{jet}} \ge 4$ & $N_{\text{jet}} \ge 4$ & $N_{\text{jet}} \ge 4$ \\
       & $N_{\text{b-tag}}\ge 1$& $N_{\text{b-tag}} = 0$ & $N_{\text{b-tag}} = 1$ & $N_{\text{b-tag}} \ge 2$ \\ \hline
\zp (M=$0.5\TeV$) & 48.5 & 14.0 & 41.1 & 34.6 \\
\zp (M=$1.0\TeV$) & 68.5 & 36.1 & 95.5 & 74.7 \\
\zp (M=$1.5\TeV$) & 56.4 & 33.9 & 76.5 & 50.9 \\
\zp (M=$2.0\TeV$) & 38.0 & 32.4 & 60.7 & 37.5 \\ \hline
\ttbar & 5612 & 2988 & 7802 & 6093 \\
$\PW$/$\Z$+jets & 1727 & 7705 & 1296 & 173 \\
Single top & 550 & 202 & 423 & 228 \\
Multijet & 164 & 195 & 104 & 152 \\
\hline
Total background & $8052 \pm 511$ & $11089 \pm 1241$  & $9626 \pm 822$ & $6646 \pm 687$ \\
Data & 8465 & 10714 & 9664 & 6697 \\
 \hline \hline
\multicolumn{5}{|c|}{Threshold analyses, electron channel}\\ \hline
Sample & $N_{\text{jet}}=3$& $N_{\text{jet}} \ge 4$ & $N_{\text{jet}} \ge 4$ & $N_{\text{jet}} \ge 4$ \\
       & $N_{\text{b-tag}}\ge 1$& $N_{\text{b-tag}} = 0$ & $N_{\text{b-tag}} = 1$ & $N_{\text{b-tag}} \ge 2$ \\ \hline
\zp (M=$0.5\TeV$) & 34.7 & 10.5 & 29.4 & 25.0 \\
\zp (M=$1.0\TeV$) & 58.9 & 32.4 & 85.0 & 67.4 \\
\zp (M=$1.5\TeV$) & 51.2 & 31.7 & 73.5 & 50.8 \\
\zp (M=$2.0\TeV$)  & 33.8 & 30.2 & 59.5 & 37.6 \\ \hline
\ttbar & 4307 & 2395 & 6183 & 4770 \\
$\PW$/$\Z$+jets & 1372 & 6355 & 1051 & 142 \\
Single top & 428 & 158 & 345 & 184 \\
Multijet & 491 & 1398 & 504 & 210 \\
\hline
Total background & $6597 \pm 442$ & $10307 \pm 1136$ & $8083 \pm 721$ & $5306 \pm 514$ \\
Data & 6932 & 10008 & 7946 & 5309 \\
 \hline
    \end{tabular}
  \end{center}
\end{table}

\clearpage

\begin{table}[htp]
\topcaption{\label{tab:eventyieldboosted} Number of expected and observed events in the boosted analyses
for an integrated luminosity of $4.4$--$5.0\fbinv$.
The narrow-width $\Zprime$ samples are normalized to cross sections times branching fractions of 1~$\rm pb$.
The other simulated samples are normalized to data as described in the text.
The uncertainty in the total background corresponds to yield changes originating from the
systematic uncertainties associated with the jet energy corrections, jet energy resolutions, b tagging, and pileup.
The normalization uncertainties on the theoretical production cross sections are summarized in 
Section 6, and are not included in the quoted value. 
The statistical uncertainties for the simulated samples are negligible.}
 \begin{center}
    \begin{tabular}{|lrrrr|} \hline
Boosted analyses & \multicolumn{2}{c}{Electron channel} & \multicolumn{2}{c|}{Muon channel} \\ \hline
       Sample & $N_{\text{b-tag}}=0$ & $N_{\text{b-tag}} \ge 1$  & $N_{\text{b-tag}}=0$ & $N_{\text{b-tag}} \ge 1$ \\
      \hline
      \zp (M$=1\TeV$)         & 17.1 & 36.5 & 27.8 & 48.3 \\
      \zp (M$=1.5\TeV$)       & 44.7 & 55.4 & 95.9 & 94.4 \\
      \zp (M$=2\TeV$)         & 62.1 & 52.8 & 146.3 & 94.1 \\
      \zp (M$=3\TeV$)         & 57.2 & 36.9 & 155.2 & 69.0 \\
      \hline
      $\rm t\bar{t}$                 & 172 &  336 &  157 &  262\\
      $\PW/Z$+jets                      &  95 &    6 &  149 &    9\\
      Single top                     &   9.3 &   15 &    8.1 &   11\\
      \hline
      Total background               & $276 \pm 58$ &  $357 \pm 50$ &  $314 \pm 72$ & $282 \pm 34$\\
      Data                      & 277   &  354   &  300   &  269 \\
      \hline
    \end{tabular}
  \end{center}
\end{table}

\section{Systematic Uncertainties}
\label{sec:systematics}

Systematic uncertainties enter the analyses
in two ways: those related to the total normalization of the
simulated samples, and those from the effects that change
both the normalization and shape of the background and expected signal distributions.

Normalization uncertainties on the theoretical production cross sections
are considered for all background processes. In some instances, larger
uncertainties are used for the boosted analyses as they probe a limited region of phase space.
The following variations on the rates, which were obtained in a previous analysis~\cite{PhysRevD.84.092004}, are included:
$\ttbar$ (15\%); single top-quark for threshold (30\%) and for boosted (50\%) analyses;
$\PW/\Z$+light-quark jets correlated (50\%) and additional Drell-Yan uncorrelated (30\%) for threshold analyses,
$\PW$+light-quark jets (50\%) and uncorrelated $\Z$+light-quark jets (100\%) for boosted analyses; $\PW/\Z$+heavy-quark jets (100\%).
In addition, a 2.2\% uncertainty in the luminosity~\cite{CMSLumi} and 3\% (5\%)
lepton trigger and identification uncertainty is applied to all simulated samples for the threshold
(boosted) analyses.

Several sources of systematic uncertainty affect both the shape and the rate
of the templates used in the analyses. The uncertainty on the
energy of jets is of the order of a few percent and is parametrized as a function of the
jet $\pt$ and $\eta$~\cite{JETJINST}. The uncertainty on the jet energy resolution varies from 6 to 20\% depending on the
jet $\eta$. The effect of both variations is propagated to the event $\MET$.
The uncertainty on the b-tagging efficiency for b jets ranges from 1.6 to 8\% depending on the jet $\eta$ and is doubled
for b jets with $\pt>670\GeV$~\cite{CMSbAlgo}. The uncertainty on the b-tagging efficiency for c jets is taken as twice the
uncertainty for b jets. The uncertainty for all other jets (mistag rate) is 11\%.

Some of the theoretical uncertainties affect the normalization and shape of the simulated samples.
A simultaneous variation of the factorization and renormalization scales to half and twice the nominal
scales is allowed for the $\ttbar$ and $\PW$+jets samples. The
matrix element to parton shower matching threshold and the amount of initial- and final-state radiation
are also allowed to vary for these samples.
A further uncertainty is included as the difference between
the $\ttbar$ production models in \textsc{powheg} and \textsc{MadGraph}.
For all simulated samples, the minimum bias cross section is varied by 1 standard deviation of its
measured value to account for the effect of pileup.

\section{Results}
\label{sec:results}

The statistical analysis is based on a binned likelihood of the
$M_{\ttbar}$ distributions in the considered channels, i.e., eight channels
for the threshold analyses and four channels for the boosted analyses.
The number of events in bin $i$ is assumed to follow a Poisson distribution with mean $\lambda_i$, given by the sum over all considered background
processes and the $\Zprime$ signal. The signal is scaled
with a signal strength modifier $\mu$, which is the signal cross section in pb:
\[
\lambda_i = \mu S_{i} + \sum_{k} B_{k,i}.
\]
Here, $k$ runs over all considered background processes,
$B_k$ is the background template for background $k$,
and $S$ is the signal template, scaled according to luminosity and a signal cross section of 1\unit{pb}.

The presence of systematic uncertainties affects the yields $\lambda_i$.
A nuisance parameter $\theta_u$ is thus introduced for each
independent source of systematic uncertainty considered.
A rate-only uncertainty is modeled with a coefficient for the template
$B_{k}$ with a log-normal prior. A rate and shape uncertainty is modeled by choosing a Gaussian prior
for $\theta_u$ and using this parameter to interpolate between the nominal template
and the shifted templates obtained by applying
a ${\pm}1\;\sigma$ systematic shift to the simulated samples.
This interpolation uses a smooth function, which is cubic in the range
${\pm}1\;\sigma$ and linear beyond ${\pm}1\;\sigma$.

We use the modified frequentist construction CL$_\mathrm{s}$~\cite{cls1,cls2} to calculate the 95\% confidence level (CL)\
upper limits on the $\Zprime \rightarrow \ttbar$ cross section. 
The expected upper limits are calculated using background-only pseudo-experiments ($\mu = 0$) and calculating
the upper limit for each pseudo-experiment. The expected limit is given by the median of the distribution of upper limits, and the central 68\% and 95\% give the $\pm1$ and $\pm2$ standard deviation (s.d.) excursions.

The number of simulated background events in the $M_{\ttbar}>2\TeV$ region that pass the boosted selection is
rather limited. To ensure a proper background modeling in the entire $M_{\ttbar}$ range, we merge bins in the
$M_{\ttbar}$ distribution requiring a minimum number of background events per bin.
The bins are chosen such that the uncertainty on the number of expected
background events due to the limited number of simulated events is not worse than 30\% in all channels.
The uncertainty due to finite size of the simulated samples is taken into account using the
``Barlow-Beeston lite'' method~\cite{Conway-PhyStat} that defines one additional nuisance parameter with a Gaussian
distribution for each bin, and performs the maximization of the likelihood with respect to these new parameters
analytically.

\begin{figure}
\centering
 \includegraphics[width=0.8\textwidth]{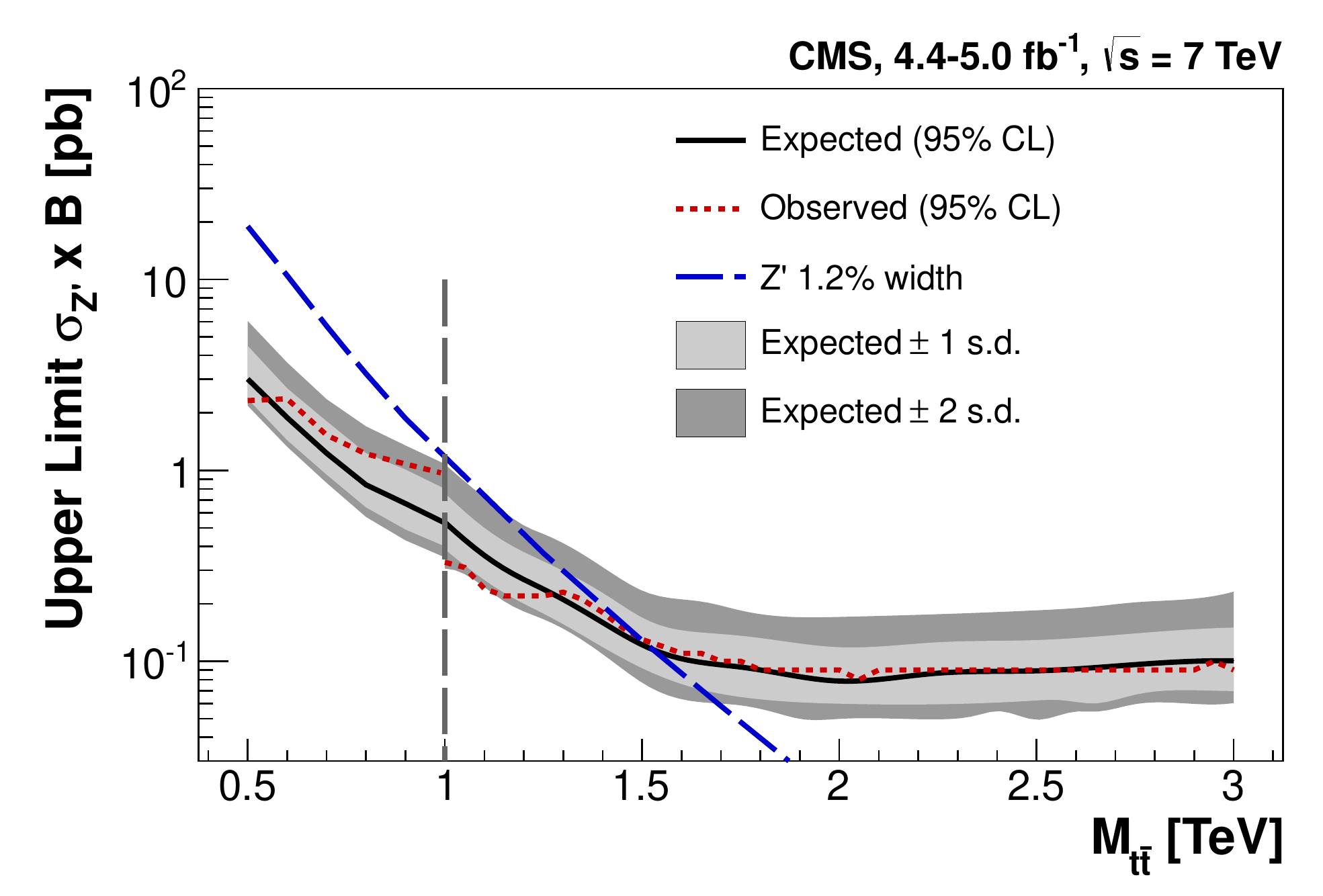}
 \includegraphics[width=0.8\textwidth]{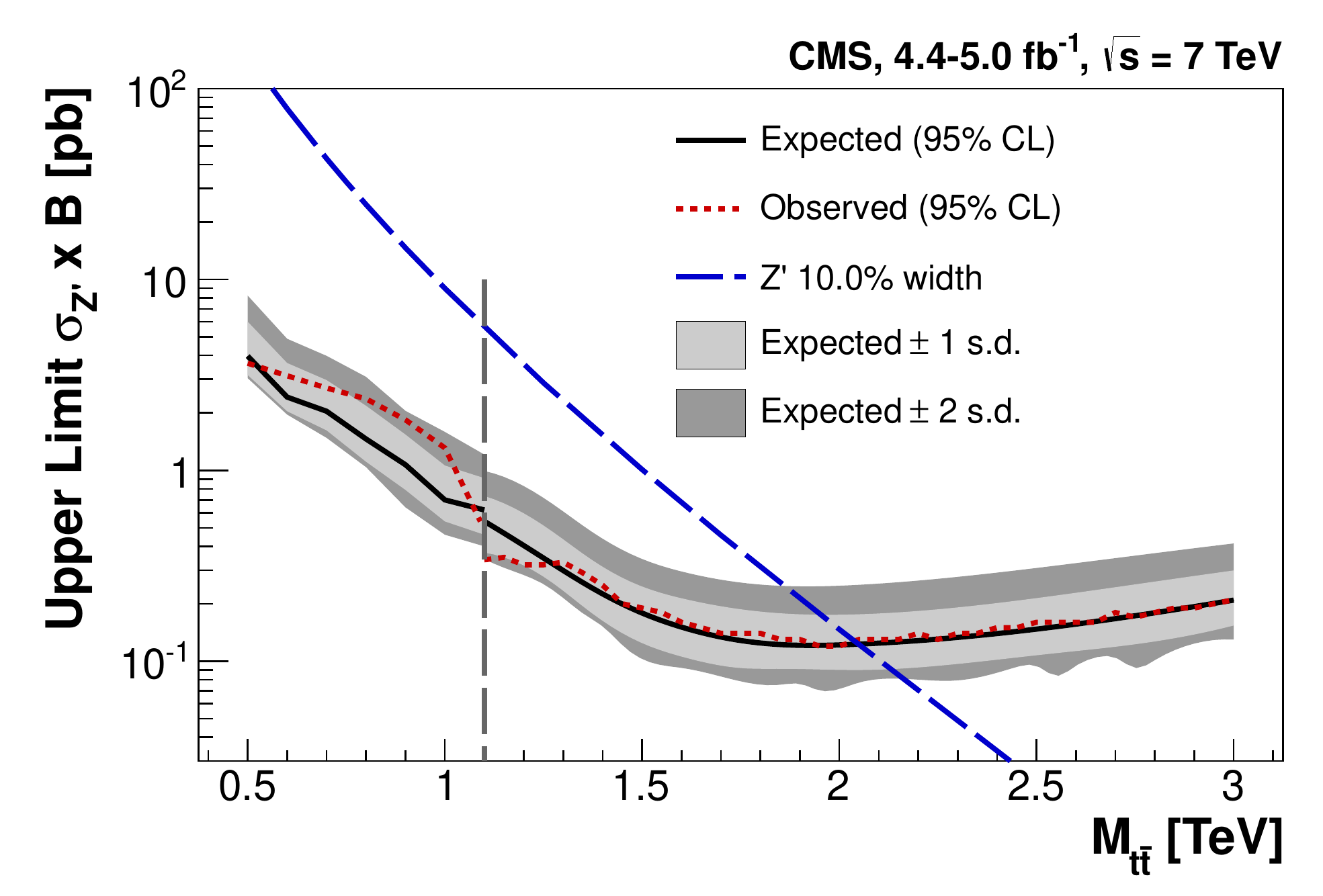}
 \caption{The 95\% CL upper limits on the product of the production cross section $\sigma_{\zp}$ and the branching fraction $B$ of
hypothesized resonances that decay into $\tt$ as a function of the invariant mass of the resonance.
The $\Zprime$  production with $\Gamma_{\Zprime}/m_{\Zprime}=$~1.2\% (a) and 10\% (b) compared to predictions based
on~\cite{Jain11124928}.  The $\pm1$ and $\pm2$ s.d.\ excursions from the expected limits are also shown. The vertical dashed line indicates the transition between the threshold and the boosted analyses, chosen based on
the sensitivity of the expected limit.}
\label{fig:limits}
\end{figure}

\begin{figure}
\centering
 \includegraphics[width=0.8\textwidth]{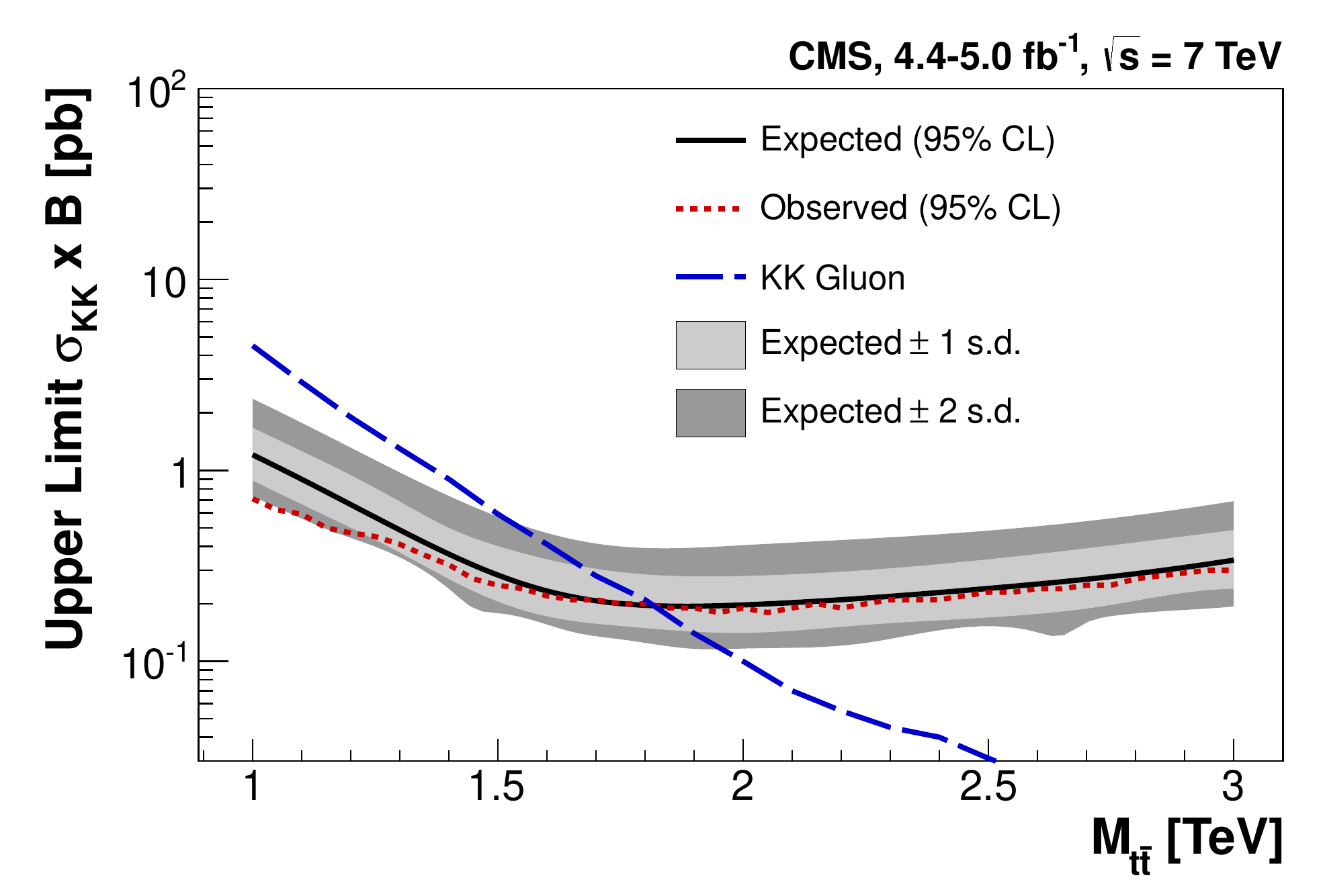}
\caption{The 95\% CL upper limits on the product of the production cross section $\sigma_\mathrm{KK}$ and the branching fraction $B$ of
Kaluza--Klein excitation of gluon production from~\cite{Agashe:2006hk}, compared to the theoretical prediction of that model. The ${\pm}1$ and ${\pm}2$ s.d.\ excursions from the expected limits are also shown. }
\label{fig:limitsKK}
\end{figure}

Figures~\ref{fig:limits} and~\ref{fig:limitsKK} show the expected and observed 95\% CL upper limits for the product
of the production cross section times branching fraction of hypothesized resonances that decay into $\tt$ as a function of the invariant mass of the resonance.
The dashed lines indicate the values predicted by various models for new physics processes.
The expected mass exclusion region for a topcolor $\Zprime$
with $\Gamma_{\Zprime}/m_{\Zprime}= 1.2$\% is
$M_{\Zprime} < 1.53\TeV$, the observed exclusion is
$M_{\Zprime} < 1.49\TeV$. For wide resonances with $\Gamma_{\Zprime}/m_{\Zprime}=10$\%,
the exclusion mass region is
$M_{\Zprime} < 2.04\TeV$ for both the expected and
observed limits. In Fig.~\ref{fig:limits}, the vertical dashed line indicates the transition between the threshold and the boosted analyses, chosen based on
the sensitivity of the expected limit. For a Kaluza--Klein excitation of a gluon ($\mathrm{g_{K K}}$)
the exclusion mass region is
$M(\mathrm{g_{K K}}) < 1.82\TeV$ for both the expected and
observed limits.

\clearpage

\section{Summary}
\label{sec:conclusions}

Results from a model-independent search for the production of heavy
resonances decaying into $\ttbar$ are presented. The data sample corresponds to an integrated
luminosity of $4.4$--$5.0\fbinv$ recorded in 2011 by the CMS detector in
$\Pp\Pp$ collisions at $\sqrt{s} = 7\TeV$ at the LHC.
After analyzing events with a lepton (muon or electron) plus jets
final state, no evidence of such massive resonances is found above the SM prediction.
Therefore, limits are set on the production of non-SM particles.
Topcolor \cPZpr\ bosons with a width of 1.2 (10)\% of the \zp mass are excluded at 95\% CL
for masses below $1.49~(2.04)\TeV$; an upper limit of $0.3~(1.3)\unit{pb}$
is set on the production cross section times branching fraction for a resonance mass of $1\TeV$.
In addition, Kaluza--Klein excitations of a gluon with masses below $1.82\TeV$ (at 95\% CL)
in the Randall--Sundrum model are excluded; an upper limit of
0.7\unit{pb} is set on the production cross section times branching fraction for a resonance mass of $1\TeV$. In both instances,
the upper limits are lower for larger resonance masses.
These results set the most stringent limits, to date, for \ttbar\ resonant production in the $0.5$--$2\TeV$ mass range.

\section*{Acknowledgments}
We congratulate our colleagues in the CERN accelerator departments for the excellent performance of the LHC machine. We thank the technical and administrative staff at CERN and other CMS institutes, and acknowledge support from: FMSR (Austria); FNRS and FWO (Belgium); CNPq, CAPES, FAPERJ, and FAPESP (Brazil); MES (Bulgaria); CERN; CAS, MoST, and NSFC (China); COLCIENCIAS (Colombia); MSES (Croatia); RPF (Cyprus); MoER, SF0690030s09 and ERDF (Estonia); Academy of Finland, MEC, and HIP (Finland); CEA and CNRS/IN2P3 (France); BMBF, DFG, and HGF (Germany); GSRT (Greece); OTKA and NKTH (Hungary); DAE and DST (India); IPM (Iran); SFI (Ireland); INFN (Italy); NRF and WCU (Korea); LAS (Lithuania); CINVESTAV, CONACYT, SEP, and UASLP-FAI (Mexico); MSI (New Zealand); PAEC (Pakistan); MSHE and NSC (Poland); FCT (Portugal); JINR (Armenia, Belarus, Georgia, Ukraine, Uzbekistan); MON, RosAtom, RAS and RFBR (Russia); MSTD (Serbia); SEIDI and CPAN (Spain); Swiss Funding Agencies (Switzerland); NSC (Taipei); TUBITAK and TAEK (Turkey); STFC (United Kingdom); DOE and NSF (USA).

\bibliography{auto_generated}   

\cleardoublepage \appendix\section{The CMS Collaboration \label{app:collab}}\begin{sloppypar}\hyphenpenalty=5000\widowpenalty=500\clubpenalty=5000\textbf{Yerevan Physics Institute,  Yerevan,  Armenia}\\*[0pt]
S.~Chatrchyan, V.~Khachatryan, A.M.~Sirunyan, A.~Tumasyan
\vskip\cmsinstskip
\textbf{Institut f\"{u}r Hochenergiephysik der OeAW,  Wien,  Austria}\\*[0pt]
W.~Adam, E.~Aguilo, T.~Bergauer, M.~Dragicevic, J.~Er\"{o}, C.~Fabjan\cmsAuthorMark{1}, M.~Friedl, R.~Fr\"{u}hwirth\cmsAuthorMark{1}, V.M.~Ghete, J.~Hammer, N.~H\"{o}rmann, J.~Hrubec, M.~Jeitler\cmsAuthorMark{1}, W.~Kiesenhofer, V.~Kn\"{u}nz, M.~Krammer\cmsAuthorMark{1}, I.~Kr\"{a}tschmer, D.~Liko, I.~Mikulec, M.~Pernicka$^{\textrm{\dag}}$, B.~Rahbaran, C.~Rohringer, H.~Rohringer, R.~Sch\"{o}fbeck, J.~Strauss, A.~Taurok, W.~Waltenberger, G.~Walzel, E.~Widl, C.-E.~Wulz\cmsAuthorMark{1}
% ...
\vskip\cmsinstskip
\dag:~Deceased\\
1:~~Also at Vienna University of Technology, Vienna, Austria\\
% ...

\end{sloppypar}
\end{document}